\documentstyle[prb,aps,12pt]{revtex}
\begin{document}
\title{Infrared conductivity of a one-dimensional charge-ordered state:
quantum lattice effects}
\author{C.A. Perroni, V. Cataudella, G. De Filippis, G. Iadonisi, V.
Marigliano Ramaglia, and F. Ventriglia}
\address{Coherentia-INFM  and Dipartimento di Scienze Fisiche, \\
Universit\`{a} degli Studi di Napoli ``Federico II'',\\
Complesso Universitario Monte Sant'Angelo,\\
Via Cintia, I-80126 Napoli, Italy}
\date{\today}
\maketitle

\begin{abstract}
The optical properties of the charge-ordering ($CO$) phase of the
one-dimensional ($1D$) half-filled spinless Holstein model are
derived at zero temperature within a well-known variational
approach improved including second-order lattice fluctuations.
Within the $CO$ phase, the static lattice distortions give rise to
the optical interband gap, that broadens as the strength of the
electron-phonon ($el-ph$) interaction increases. The lattice
fluctuation effects induce a long subgap tail in the infrared
conductivity and a wide band above the gap energy. The first term
is due to the multi-phonon emission by the charge carriers, the
second to the interband transitions accompanied by the
multi-phonon scattering. The results show a good agreement with
experimental spectra.
\end{abstract}

\newpage

In the last years there has been a renewed interest in charge density wave
($CDW$) materials. \cite{gruner} The transition to a $CO$ phase is
common to a wide range of compounds, \cite{springer} including many
quasi-$1D$ materials such as organic conjugated polymers, charge
transfer salts, molybdenum bronzes and $MX$ chains.
\cite{gruner,springer,kago,travi,testa}
These materials undergo a Peierls instability driven by the $el-ph$
interaction in the half-filled band case. For most quasi-$1D$ systems,
the lattice zero-point motion is comparable to the Peierls lattice
distortion, so quantum lattice fluctuations must be taken into account
to satisfactorily describe spectral, transport and optical properties.
\cite{ross,wilkins,perfetti}
In particular the study of the optical absorption of $CO$ materials
represents a very useful tool to extract the gap energy and, in
general, to investigate the properties of the ordered state.
\cite{gruner,gruner1,degiorgi,vescoli}
In this framework, the experimental measurements have pointed out that
the lattice fluctuation effects can remove the inverse square-root
singularity expected for the case of a static distorted lattice
\cite{anderson} affecting profoundly the conductivity spectra.
\cite{ross,degiorgi1,schwartz}
Actually these effects give rise to the subgap optical absorption seen
in these materials where a significant tail below the maximum of the
interband transition term is measured. Moreover the optical absorption
above the interband optical gap band also presents deviations from the
behavior obtained within the mean-field approach of the static
lattice. \cite{degiorgi1}

The challenge of understanding the effect of quantum lattice
fluctuations on the Peierls dimerization and the absorption
spectra has determined an intense study of the Holstein model
\cite{holstein} that is a typical $el-ph$ coupling model
developing a $CO$ state at half-filling. Actually the Holstein
Hamiltonian has been investigated by using various techniques,
such as Monte-Carlo simulations, \cite{hirsch,ross1}
renormalization-group analysis, \cite{bourb,schme} variational
method, \cite{zheng,zhe1,wang} density-matrix renormalization
group \cite{bursill} and exact diagonalization. \cite{weisse}
These studies reveal that, in the spinless case, there is a
quantum phase transition from a Luttinger liquid (metallic) phase
to an insulating phase with $CDW$ long-range order. Because of
limited system sizes in numerical approaches, except for the
antiadiabatic regime, the behavior of the conductivity spectra is
not well determined, so the extraction of the gap from the optical
data is not precise. \cite{weisse} However, through exact
diagonalization methods, the spectral weight of the conductivity
can be deduced showing the onset of the infrared absorption for
lattices of a few sites. Recently, an analytical variational
approach, \cite{zhe1,wang} valid in the weak $el-ph$ coupling
regime, has been developed to study the phase diagram and the
optical conductivity of the Holstein model. The employed
approximation has different effects on the phase diagram and the
optical properties. In fact, the maximum in the optical spectra is
not directly corresponding to the gap calculated within the
variational approach. Actually the peak position of the optical
conductivity is higher than the gap with an energy separation of
the order of the electron transfer integral in the adiabatic
limit. Therefore, experimental findings of a tail in the optical
spectra below the gap energy (corresponding to the maximum of
experimental conductivities) does not find a clear explanation
within previous approaches.

In this paper we employ the variational scheme proposed for the
$1D$ half-filled spinless Holstein model by H. Zheng, D. Feinberg
and M. Avignon \cite{zheng} ($ZFA$) in order to calculate the
spectral properties and the infrared response. We note that the
$ZFA$ method improves the mean-field solution showing a partial
agreement with other numerical works
\cite{hirsch,ross1,bursill,weisse}. Actually this approach is able
to introduce lattice fluctuations on the mean-field Peierls
solution since it takes into account the nonadiabatic polaron
formation. Indeed, due to the polaronic effect, the lattice
deformation is allowed to follow instantaneously the electrons.
The calculated conductivity spectra are characterized by a
transfer of spectral weight from low to high energies and by a
broadening of the optical interband gap, with increasing the
$el-ph$ coupling. The effect of the quantum lattice fluctuations
is able to determine in the infrared conductivity a subgap
absorption term near the phonon energy and a wide band above the
gap energy. The first contribution is due to the multi-phonon
emission by the charge carriers, the second to the interband
transitions accompanied by the multi-phonon scattering. The
inclusion of lattice fluctuation effects beyond the $ZFA$ approach
is able to smooth the inverse square-root singularity of the $ZFA$
and mean-field conductivity. Moreover these effects strengthen the
features of the conductivity below and above the gap already found
in the $ZFA$ approach inducing an actual subgap tail. Therefore
lattice fluctuations influence the optical absorption at the
low-energy subgap, at the gap and at the high-energy scale above
the interband optical gap term. These features are found in the
measured spectra.\cite{degiorgi1,schwartz}

In section I the model and the $ZFA$ variational approach are
briefly reviewed; in section II the spectral properties are
deduced in order to characterize the quasi-particle gap of the
$CO$ phase; in section III the infrared spectra obtained within
the $ZFA$ approach are discussed; finally in section IV the
effects of lattice fluctuations beyond the $ZFA$ approach are
analyzed.

\section{Model and ZFA variational approach}
In this paper we deal with the $1D$ spinless Holstein model \cite{holstein}
at half-filling.

The Holstein Hamiltonian is

\begin{eqnarray}
H=-t \sum_{<i,j>} c^{\dagger}_{i} c_{j}
 +\omega_0 \sum_{i} a^{\dagger}_{i}a_{i}
+g \omega_0 \sum_{i} c^{\dagger}_{i}c_{i} \left(
a_{i}+a^{\dagger}_{i} \right) - \mu \sum_{i} c^{\dagger}_{i} c_{i}
.  \label{1r}
\end{eqnarray}
Here  $t$ is the electron transfer integral between nearest
neighbor ($nn$) sites $<i,j>$, $c^{\dagger}_{i} \left( c_{i}
\right)$ creates (destroys) an electron at the $i$-th site  and
$\mu$ is the chemical potential. In the second term of
eq.(\ref{1r}) $ a^{\dagger}_{i} \left( a_{i} \right)$ is the
creation (annihilation) phonon operator at the site i, $\omega_0$
denotes the frequency of the optical phonon mode and the parameter
$g$ represents the coupling constant between electrons and
phonons. The dimensionless parameter $\lambda$
\begin{equation}
\lambda= \frac{g^2 \omega_0}{2 t},
\label{1r1b}
\end{equation}
indicating the ratio between the small polaron binding energy and the energy
gain of an itinerant electron on a rigid lattice, is useful to measure the
strength of the $el-ph$ interaction in the adiabatic regime.
We consider spinless electrons since they, even if at a very rough level,
mimic the action of an on-site Coulomb repulsion preventing the formation of
local pairs. Actually, for one dimensional systems in the limit of
infinite local repulsion $U$, the charge sector of the Hubbard model
maps onto a spinless model, therefore the spinless Holstein model can
be considered as a reliable model for typical one-dimensional systems.

The hopping of electrons is supposed to take place between the equivalent
$nn$ sites of a $1D$ lattice separated by the distance $a$.
The units are such that the Planck constant $\hbar=1$ and the lattice
parameter $a$=1.

As stressed in the original $ZFA$ paper, \cite{zheng} the starting
point of the approach is the consideration that the strong
coupling and infinite phonon frequency limit of the model are
described by polarons. The $ZFA$ approach extends the polaron
formation to intermediate regimes recovering the mean-field
solution of the zero phonon frequency limit. Following the $ZFA$
variational scheme, three successive canonical transformations are
performed in order to treat the electron-phonon interaction
variationally and to introduce the charge-ordering solution.

The first transformation is the variational Lang-Firsov unitary one \cite{lang}
\begin{equation}
U_{1}=\exp \left[ g \sum_{j}
\left( f c^{\dagger}_{j} c_{j} +\Delta_{j} \right)
\left( a_{j}-a^{\dagger}_{j} \right)
\right],
\label{2r}
\end{equation}
where $f$ and $\Delta_{j}$ are variational parameters. The
quantity  $f$ controls the degree of the antiadiabatic polaronic
effect since the lattice deformation is allowed to follow
instantaneously the electrons. Moreover $\Delta_{j}$ denotes a
displacement field describing lattice distortions due to the
average electron motion. At half-filling the charge-ordered
solution is obtained by assuming
\begin{equation}
\Delta_{i}=\Delta+\Delta_{CO} e^{i Q R_{i} },
\label{8ra}
\end{equation}
where $\Delta$ represents the lattice distortion unaffected by the
instantaneous position of electrons and $\Delta_{CO}$ the additional local
lattice distortion due to the Peierls dimerization with $ Q= \pi$.

The second transformation is
\begin{equation}
U_{2}=\exp \left[\alpha \sum_{j} \left( a^{\dagger}_{j}
a^{\dagger}_{j} -a_{j}a_{j} \right)\right],
\label{3r}
\end{equation}
where the variational parameter $\alpha$ determines a phonon frequency
renormalization.

The transformed Hamiltonian
$\tilde{H}=  U_{2}^{-1} U_{1}^{-1} H U_{1} U_{2} $ is
\begin{eqnarray}
\tilde{H}=&&
-t \sum_{<i,j>} X^{\dagger}_i X_j  c^{\dagger}_{i} c_{j}
+\bar{\omega}_0 \sum_{i} a^{\dagger}_{i}a_{i}+
L \omega_0 \sinh^2\left(2\alpha\right)
+  g^{2} \omega_{0} \sum_{i} \Delta_{i}^{2}    \nonumber \\
&&
+\omega_0 \sinh\left(2 \alpha\right) \cosh\left( 2 \alpha\right)
\sum_{i} \left( a^{\dagger}_{i}a^{\dagger}_{i} +a_{i}a_{i} \right)
-g \omega_{0} e^{2 \alpha}
\sum_{i} \Delta_{i} \left( a_{i} +a^{\dagger}_{i} \right) \nonumber \\
&&
+g \omega_{0} \left( 1-f \right) e^{2 \alpha}
\sum_{i} c^{\dagger}_{i}c_{i} \left( a_{i} +a^{\dagger}_{i} \right)
+  \sum_{i} c^{\dagger}_{i}c_{i} \left( \eta_{i} -\mu \right),
\label{4r}
\end{eqnarray}
where we have introduced the phonon operator  $X_i$

\begin{equation}
X_i=\exp \left[ g f e^{-2 \alpha} \left( a_{i} - a^{\dagger}_{i} \right)
\right],
\label{6r}
\end{equation}
the renormalized phonon frequency
$ \bar{\omega}_0=\omega_0 \cosh(4 \alpha)$, the number of  lattice
sites $L$
and the quantity $\eta_{i}$

\begin{equation}
\eta_{i} = g^{2} \omega_{0}  f \left( f-2 \right)+2 g^{2} \omega_{0}
\left( f-1 \right) \Delta_{i}.
\label{8r}
\end{equation}

In the $ZFA$ approach the energy is deduced introducing a test Hamiltonian
characterized by non interacting electron and phonon degrees of freedom such
that $ \langle \tilde{H}-H_{test} \rangle _{t}=0 $, where
$ <>_t $ indicates the mean value obtained by using the ground state of
$H_{test}$.
The test Hamiltonian is given by

\begin{eqnarray}
H_{test} &=&
-t_{eff} \sum_{<i,j>} c^{\dagger}_{i}c_{j} +
\bar{\omega}_0 \sum_{i}a^{\dagger}_{i}a_{i}+
L \omega_0  \sinh^2\left( 2 \alpha\right)
+L g^{2} \omega_{0}  \left( \Delta^{2}+\Delta^{2}_{CO} \right)   \nonumber \\
&&
-2 g^{2} \omega_{0} \Delta_{CO} \left( 1-f \right)
\sum_{i} e^{i Q  R_{i} } c^{\dagger}_{i}c_{i}
-\mu_{0} \sum_{i} c^{\dagger}_{i}c_{i},
\label{9r}
\end{eqnarray}
where the subsidiary chemical potential $\mu_{0}$ is
\begin{equation}
\mu_{0} =\mu-g^{2} \omega_{0}  f \left( f-2 \right)+2 g^{2} \omega_{0}
\left( f-1 \right) \Delta.
\end{equation}
The quantity $t_{eff}= t  e^{-S}$ denotes the effective transfer
integral, where the quantity
\begin{equation}
S=g^{2} f^{2} e^{-4 \alpha}
\label{9rz}
\end{equation}
controls the band renormalization due to the nonadiabatic polaron effect.
The electronic part of the test Hamiltonian is diagonalized by a third
canonical Bogoliubov transformation \cite{zheng} yielding
\begin{eqnarray}
\tilde{H}_{test} &=&
\sum_{k \epsilon NZ}
\left( \xi_{k}^{(+)}  -\mu_{0} \right) d^{\dagger}_{k} d_{k} +
\sum_{k \epsilon NZ}
\left( \xi_{k}^{(-)}  -\mu_{0} \right)  p^{\dagger}_{k} p_{k}+
\bar{\omega}_0 \sum_{q} a^{\dagger}_{q}a_{q}+
\nonumber \\
&&
+L \omega_0  \sinh^2\left( 2 \alpha\right)
+L g^{2} \omega_{0} \left( \Delta^{2}+\Delta^{2}_{CO} \right),
\label{9rb}
\end{eqnarray}
where $d^{\dagger}_{k} \left( d_{k} \right)$ creates (destroys)
a quasi-particle in the  upper band
$ \xi_{k}^{(+)} = \sqrt{ \tilde{\epsilon}^{2}_{k}+E^{2} } $ ,
 $p^{\dagger}_{k} \left( p_{k} \right)$ creates (destroys)
a quasi-particle in the lower band $ \xi_{k}^{(-)} = -\sqrt{
\tilde{\epsilon}^{2}_{k}+E^{2} } $ and $\tilde{\epsilon}_{k}$ is
the polaronic band. We note that $NZ$ indicates the New
(Brillouin) Zone defined by the condition $\tilde{\epsilon}_{k}
\leq  0$. In the $CO$ phase a gap opens between the upper and
lower bands in the quasi-particle spectrum and it is twice the
quantity
\begin{equation}
E= 2  g^{2} \omega_{0} \Delta_{CO} \left( 1-f \right) .
\end{equation}

Within the variational approach the kinetic energy mean value
$\bar{E}_{kin}$ is
\begin{equation}
\bar{E}_{kin}=<\hat{T}>= -\int_{- \tilde{W}}^{0} d \epsilon
g(\epsilon) \frac { \epsilon^{2} }{\sqrt{\epsilon^2+E^2 }  },
\label{562ra}
\end{equation}
where $\tilde{W}= 2 t_{eff}$ is the renormalized band half-width,
$g \left( \epsilon \right)$ the $1D$ density of states, and the
electron order parameter $m_e$ is given by
\begin{equation}
m_e= \left( \frac{1}{L} \right) \sum_{i} e^{i \vec{Q} \cdot
\vec{R_{i}} } \langle c^{\dagger}_{i}c_{i} \rangle = E
\int_{-\tilde{W}}^{0} d \epsilon \frac { g(\epsilon) } {
\sqrt{\epsilon^2+E^2 }  }. \label{562rb}
\end{equation}
The $CO$ phase is characterized by the order parameter different
from zero.

In Fig.1 we report the phase diagram \cite{zheng} derived within
the $ZFA$ approach in the thermodynamic limit. The ordered state
is separated with a transition line by the $A$ phase that
represents the disordered phase ($\Delta_{CO}=0$). Previous
studies \cite{ross1,zhe1,bursill,weisse} have pointed out that the
normal state has the properties of  a Luttinger liquid. This has
been verified also by using the $ZFA$ wave-function and making a
finite-size scaling analysis. \cite{weisse} In the inset of Fig.1,
there is the comparison between the transition lines calculated in
a mean-field approach (white squares), $ZFA$ approach (white
circles) and $DMRG$ approach \cite{bursill} (black squares). Here
mean-field approach means that we are neglecting the effect of
polaron formation ($f=0$, $\alpha=0$). In the ranges of parameters
relevant for quasi one-dimensional materials, the ZFA approach
improves the mean-field solution since the $CO$ phase is stable
for larger values of the $el-ph$ couplings.  However the lattice
fluctuation effects introduced by this approach are not sufficient
to obtain transition lines comparable with those of $DMRG$
approach. This indicates that quantum fluctuation effects beyond
the $ZFA$ scheme are not negligible and can play a role also in
the calculation of the dynamical properties of the model.

\section{Spectral properties within ZFA approach}

In this section we calculate the spectral properties within the $ZFA$
approach. They are discussed in order to characterize the gap in the
quasi-particle spectrum.

The electron retarded Green's function can be disentangled into electronic
and phononic terms \cite{perroni} by using the test Hamiltonian (\ref{9r}),
hence

\begin{eqnarray}
G_{ret} (k,t)= &&  e^{-S} G_{ret}^{(co)}(k,t)+ e^{-S} \left[ \exp
\left( S e^{-i \bar{\omega}_0 t} \right) -1 \right] \frac{1}{L}
\sum_{k_1 \epsilon NZ} G_{ret}^{(+)}(k_1,t)+
\nonumber \\
&& e^{-S} \left[ \exp \left( S e^{i \bar{\omega}_0 t} \right) -1
\right] \frac{1}{L} \sum_{k_1 \epsilon NZ}  G_{ret}^{(-)}(k_1,t).
\label{16ar}
\end{eqnarray}
In equation (\ref{16ar}) $G_{ret}^{(co)}(k,t)$ is
\begin{equation}
G_{ret}^{(co)}(k,t) =u^2_{k} G_{ret}^{(+)}(k,t)+ v^2_{k}
G_{ret}^{(-)}(k,t),
\label{16arzo}
\end{equation}
where $G_{ret}^{(+)}(k,t)$ and $G_{ret}^{(-)}(k,t)$ are the
Green's functions associated to the quasi-particles of the upper
and the lower bands, respectively, with $u^2_{k}$ given by
\begin{equation}
u^2_{k}=
\frac {1} {2}
\left[
1+ \frac {\tilde{\epsilon}_{k} }
{ \sqrt{\tilde{\epsilon}_{k}^2+E^2 }  }
\right]
\label{17rb}
\end{equation}
and  $v^2_{k}$ by
\begin{equation}
v^2_{k}=
\frac {1} {2}
\left[
1 - \frac {\tilde{\epsilon}_{k} }
{ \sqrt{\tilde{\epsilon}_{k}^2+E^2 }  }
\right].
\label{17rc}
\end{equation}
One obtains the spectral function
\begin{eqnarray}
A(k, \omega) = -2 \Im G_{ret} ( k, \omega) = &&
2 \pi e^{-S}
\left[
u^2_{k} \delta \left( \omega - \xi_{k}^{(+)} \right)+
v^2_{k} \delta \left( \omega - \xi_{k}^{(+)} \right)
\right]+
\nonumber \\
&&  2 \pi e^{-S}  \sum_{n=0}^{\infty}
\frac{S^n}{n!}
\left[
H(\omega - n \bar{\omega}_0) + H(-\omega - n \bar{\omega}_0)
\right],
\label{21r}
\end{eqnarray}
where the function $H(\omega)$ is
\begin{equation}
H(\omega)=\frac { g\left( \sqrt{\omega^2-E^2} \right) }
{\sqrt{1-\frac{E^2}{\omega^2}}}
\theta \left(\omega-E \right) \theta \left(\sqrt{E^2+W^2}-\omega \right),
\label{16re}
\end{equation}
with $\theta(x)$ Heaviside function. Two physically distinct terms
\cite{perroni} appear in eq.(\ref{21r}): the coherent and
incoherent one. The first derives from the coherent motion of
charge carriers and their surrounding phonon cloud. In the normal
phase it represents the purely polaronic band contribution and
shows a delta behavior. In the $CO$ phase, this term is equal to
the result of the mean-field approach when one neglects the
renormalization of the upper and lower band due to the polaron
effect. The incoherent term in eq.(\ref{21r}) is due to processes
changing  the number of phonons during the hopping of the charges.
and provides a contribution spreading over a wide energy range.

In Fig.2 we report the renormalized density of states $ N(\omega)$
calculated in the $ZFA$ approach (solid line) and mean-field
(dashed line) at a fixed value of the $el-ph$ coupling and
$t=5\omega_0$. In the $CO$ phase, a gap opens in the
quasi-particle spectrum and it is larger for the mean-field
solution than the $ZFA$ one. We note that, at the energies
corresponding to the gap, the inverse square-root singularity
occurs for both approaches. The other sharp feature in the density
of states derived in the $ZFA$ approach is due to one-phonon
processes in the upper and lower bands that are relevant in the
intermediate $el-ph$ coupling regime.

\section{Optical properties within ZFA approach}

In this section we focus our attention on the optical properties
within the $ZFA$ approach.
Since we are primarily interested to the absorption spectra in the $CO$ phase,
we evaluate the conductivity for the frequency $\omega$ different from zero.

In a regime of linear response the real part of the conductivity is
given by the current-current correlation function
\begin{equation}
\Re \sigma (\omega)=
\lim_{\beta \rightarrow \infty }
\left(\frac{1-e^{-\beta \omega}}{2 \omega L} \right)
\int_{- \infty} ^{\infty} e^{i \omega t}
\langle j^{\dagger} (t) j (0) \rangle,
\label{54r}
\end{equation}
where $\beta$ is the inverse of the temperature and $j$ the current
operator.
Performing the two canonical transformations (\ref{2r},\ref{3r}) and
making the decoupling\cite{perroni} of the correlation function in the
electron and phonon terms through the introduction of $H_{test}$ (\ref{9r}),
we get

\begin{equation}
\langle j^{\dagger} (t) j (0) \rangle=
\sum_{i,\delta} \sum_{i^{\prime},\delta^{\prime}}
( \delta \cdot \delta^{\prime} )
\Phi (i,i^{\prime},\delta,\delta^{\prime},t)
\Delta (i,i^{\prime},\delta,\delta^{\prime},t),
\label{57r}
\end{equation}
where the function
$\Delta\left(i,i^{\prime},\delta,\delta^{\prime}, t \right)$
denotes the electron correlation function
\begin{equation}
\Delta\left( i,i^{\prime},\delta,\delta^{\prime}, t \right)=
\langle c_i^{\dagger}(t) c_{i+\delta}(t)
c_{i^{\prime}+\delta^{\prime}}^{\dagger} c_{i^{\prime}} \rangle_t
\end{equation}
and the function $\Phi\left( i,i^{\prime},\delta,\delta^{\prime},t
\right)$ the phonon correlation function

\begin{equation}
\Phi\left( i,i^{\prime},\delta,\delta^{\prime} ,t \right) =
\langle X_i^{\dagger}(t) X_{i+\delta}(t)
X_{i^{\prime}+\delta^{\prime}}^{\dagger} X_{i^{\prime}} \rangle_t.
\label{59r}
\end{equation}
In order to simplify the analysis of our results, we separate $\Phi$ into two
contributions
\begin{eqnarray}
\Phi\left( i,i^{\prime},\delta,\delta^{\prime},t \right) & = &
\left[ \langle X_i^{\dagger} X_{i+\delta \hat{\alpha}}\rangle_t \right]^2
\left\{
\Phi\left( i,i^{\prime},\delta,\delta^{\prime},t \right) -
\left[ \langle X_i^{\dagger} X_{i+\delta \hat{\alpha}}\rangle_t \right]^2
\right\} \nonumber \\
&=&
 e^{ -2 S} +
\left[ \Phi\left( i,i^{\prime},\delta,\delta^{\prime},t \right)-
e^{ -2 S} \right], \label{63r}
\end{eqnarray}
where $S$ is given by eq.(\ref{9rz}).

Considering eq.(\ref{63r}), the conductivity can be expressed as a
sum of two terms \cite{perroni}

\begin{equation}
\Re \sigma(\omega)=
\Re \sigma^{(coh)}(\omega)
+ \Re \sigma^{(incoh)}(\omega).
\label{63ra}
\end{equation}
As in the spectral properties, the appearance of two physically
distinct contributions, the coherent and incoherent one, occurs.
Actually the first term $\Re \sigma^{(coh)}$ is due to the charge
transfer affected by the interactions with the lattice but not
accompanied by processes changing the number of phonons. On the
other hand, the incoherent term $ \Re \sigma^{(incoh)}$ derives
from inelastic scattering processes of emission and absorption of
phonons. Both terms of the conductivity can be expressed in terms
of the Green's function $G_{ret}^{(co)}(k,t)$ given in
eq.(\ref{16arzo}).

The coherent conductivity is derived as

\begin{equation}
\Re \sigma^{(coh)}(\omega)=
\left( \frac{ 4 \pi e^2 t^2}{\omega} \right) E^2 e^{ -2 S}
\int_{-\tilde{W} }^{0} d \epsilon
 \frac{g(\epsilon)}{(\xi^{(+)})^2}
\left(1-  \frac{\epsilon^2} {\tilde{W}^2} \right) \delta \left(
\omega-2 \xi^{(+)}  \right),
 \label{130ar}
\end{equation}
with $\xi^{(+)}=\sqrt{\epsilon^2+E^2}$. We note that this term
gives contribution to the conductivity only in the $CO$ phase
since it directly depends on the semi-gap $E$. Furthermore it is
equal to the mean-field conductivity \cite{anderson} when the
renormalization of the upper and lower bands due to the polaron
effect is neglected.

The incoherent term of the conductivity can be divided into two
components:
\begin{equation}
\Re \sigma^{(incoh)}(\omega)=
\Re \sigma^{(incoh)}_1(\omega)
+ \Re \sigma^{(incoh)}_2(\omega).
\label{63rb}
\end{equation}
The quantity $\Re \sigma^{(incoh)}_1(\omega)$ is due to the
multi-phonon emission by the charge carriers in the lower band
$\xi_{k}^{(-)}$ that does not change the electron momentum. This
first term reads
\begin{equation}
\Re \sigma^{(incoh)}_1(\omega)=
\frac{ 2 \pi e^2 t^2}{\omega} e^{ -2 S}
 \left[
\frac{1}{L}  \sum_{k \epsilon NZ} \cos (k)
(v_k^2-u_k^2)
\right] ^2
 \sum_{n=1}^{\infty} \frac{(2S)^n}{n!}
\delta (\omega - n \bar{\omega}_0),
\end{equation}
that, making the envelope of the delta functions (procedure exact in the limit
$\omega_0 \rightarrow 0$), becomes
\begin{equation}
\Re \sigma^{(incoh)}_1(\omega)=
\frac{ \pi e^2 }{2 \omega \bar{\omega}_0}  \bar{E}_{kin}^2
\frac{ (2S)^{\omega / \bar{\omega}_0} }
{ \Gamma(1+ \omega / \bar{\omega}_0 ) }
\theta (\omega - \bar{\omega}_0 ),
\end{equation}
where $\bar{E}_{kin}$ is the mean value of the kinetic energy equal to
eq.(\ref{562ra}) and $\Gamma(x)$ is the gamma function.

In eq.(\ref{63rb}), $\Re \sigma^{(incoh)}_2(\omega)$ takes into
account the interband transitions accompanied by multi-phonon
scattering. This second term is given by
\begin{equation}
\Re \sigma^{(incoh)}_2(\omega)=
\frac{ 2 \pi e^2 t^2}{\omega} e^{ -2 S}
\frac{1}{L}  \sum_{k_1,k_2 \epsilon NZ}
(1+ 4 v_{k_1} u_{k_1}  v_{k_2} u_{k_2})
 \sum_{n=1}^{\infty} \frac{(2S)^n}{n!}
\delta (\omega - n \bar{\omega}_0 +\xi_{k_1}^{(-)} - \xi_{k_2}^{(+)}),
\end{equation}
that, enveloping the delta functions, can be transformed as

\begin{eqnarray}
\Re \sigma^{(incoh)}_2(\omega)= &&
\left( \frac{ 2 \pi e^2 t^2}{\omega \bar{\omega}_0 } \right) e^{ -2 S}
\int_{-\tilde{T_1} }^{0} d \epsilon_1
\int_{-\tilde{T_2} }^{0} d \epsilon_2
\frac{ (2S)^y } { \Gamma(1+y) }
g(\epsilon_1) g(\epsilon_2)
\left( 1 +  \frac{E^2}{4 \xi^{(+)}_1 \xi^{(+)}_2 }
\right)
\times
\nonumber \\
&&
\times
\theta( \omega - \bar{\omega}_0 -2 E)
\label{134ar}
\end{eqnarray}
where $T_1= \min{\left(\tilde{W},\sqrt{( \omega - \bar{\omega}_0 -E)^2-E^2}
\right)}$,
 $T_2= \min{\left(\tilde{W},
\sqrt{( \omega - \bar{\omega}_0 -\xi_1^{(+)})^2-E^2}\right)}$,
$\xi_{i}^{(+)}$ is
defined as
\begin{equation}
\xi_{i}^{(+)}=\sqrt{\epsilon^2_{i}+E^2},
\end{equation}
with $i=1,2$, and $y=(\omega-\xi^{(+)}_1-\xi^{(+)}_2 )/\bar{\omega}_0$.

We have checked that the sum rule
\begin{equation}
\int_0^{\infty} d\omega \Re \sigma(\omega)=
- \frac{\pi}{2} e^2 \bar{E}_{kin}
\label{560r}
\end{equation}
is verified by the calculated conductivity spectra in the $CO$
phase, where $\bar{E}_{kin}$ is given by eq.(\ref{562ra}).

In Fig.3a we report the different contributions to the
conductivity spectrum derived within the $ZFA$ approach for $t=5
 \omega_0$ and $g=2.4$: the coherent term (dashed line), the
incoherent term due to multi-phonon emissions (dotted line) and
the incoherent term in correspondence with interband transitions
(dash-dotted line). For this value of the $el-ph$ coupling, the
gap is larger than the phonon frequency, that represents the
absorption threshold of the first incoherent term. The incoherent
term at $\omega_0$ reaches the largest values in the intermediate
coupling regime where the nonadiabatic polaron formation is more
effective. This contribution is still present for large $el-ph$
couplings giving rise to a subgap absorption band. The second term
of the incoherent conductivity spreads for a wide range of
frequencies above the energy gap following the coherent interband
absorption band.

In Fig.3b the conductivity spectra within $ZFA$ (solid line) and
mean-field (dashed line) approach at $t=5  \omega_0$ and $g=2.4$
are compared. Not only the optical gap within $ZFA$ is smaller
than mean-field one, but the two incoherent terms due to lattice
fluctuation effects are able to provide a non negligible
contribution below and above the gap (the arrow in figure
indicates this last contribution). This suggests that a better
treatment of the lattice fluctuations allows to capture features
of the optical conductivities that are found in experimental
spectra. \cite{degiorgi1} In the next section we will see that
lattice fluctuation effects beyond the $ZFA$ approach give rise in
the conductivity to an actual subgap tail in good agreement with
experimental data.

With rising the $el-ph$ coupling, a transfer of spectral weight
from low to high energies takes place and the optical gap
broadens. The optical response within $ZFA$ approach is strongly
dependent on adiabaticity ratio since the incoherent terms acquire
increasing spectral weight compared with that of the coherent
interband term as the ratio $t/\omega_0$ decreases. Actually the
spectral weight can be measured through the $\omega$-integrated
function
\begin{equation}
S(\omega)= \int_0^{\omega} d\omega^{\prime} \Re
\sigma(\omega^{\prime}),\label{560rg}
\end{equation}
whose value $S_m$ in the limit of infinite frequency is given by
eq.(\ref{560r}).

\section{Fluctuations beyond ZFA approach}
In this section we deal with quantum lattice fluctuation effects
beyond the $ZFA$ approach. We first determine the scattering rate
of the quasi-particles of the upper and lower bands. Next we will
analyze the effect of the self-energy insertions on the infrared
conductivity that is the main aim of this paper. This is performed
following the lines of our previous works. \cite{perroni,perroni1}

Now we consider the actual transformed Hamiltonian of the system
in eq.(\ref{4r}) as a perturbation of the test Hamiltonian in
eq.(\ref{9r}). At the second order of the perturbation theory the
retarded self-energy $\Sigma_{ret}^{(2)} \left(k, \omega\right)$
can be derived. \cite{schna,loos1,loos2,kada,fehske} If only the
dominant autocorrelation terms are retained, the self-energy is
local and provides the scattering rates of the quasi-particles of
the upper and lower bands
\begin{equation}
\Gamma_{\pm}( {k})=-2 \Im \Sigma^{(2)}_{ret} \left( \omega=
\xi^{(\pm)}_{{k}} \right). \label{z42r}
\end{equation}
These two quantities turn out to be equal, so we have only one
scattering rate
\begin{equation}
\Gamma( {k})=\Gamma_{+}( {k})=\Gamma_{-}( {k}).
\end{equation}
The scattering rate can be decomposed as
\begin{equation}
\Gamma(k)=\Gamma(\xi^{(+)}_{k})= \Gamma_{1-phon}(\xi^{(+)}_{k})+
\Gamma_{multi-phon}(\xi^{(+)}_{k}),
\label{44r}
\end{equation}
where $\Gamma_{1-phon}$ is the contribution due to single phonon
processes only

\begin{eqnarray}
\Gamma_{1-phon}(\xi^{(+)}_{k})=&& 2 Z t^2 e^{ -2 S} g^2 f^2 e^{-4
\alpha}  g_{1,l=1}(\xi^{(+)}_{k})+ g^2 \omega_0^2 e^{ 4 \alpha }
(1-f)^2 g_2(\xi^{(+)}_{k}), \label{45r}
\end{eqnarray}
$\Gamma_{multi-phon}$ represents the scattering rate by
multiphonon processes

\begin{equation}
\Gamma_{multi-phon}(\xi^{(+)}_{k})= Z t^2 e^{ -2 S} \sum_{l=2}^{+
\infty} \frac{\left( 2 g^2 f^2 e^{-4 \alpha} \right)^l }{l!}
g_{1,l}(\xi^{(+)}_{k}). \label{46r}
\end{equation}
In the previous equations the function $g_{1,l}(\xi^{(+)}_{k})$
reads

\begin{eqnarray}
g_{1,l}(\xi^{(+)}_{k}) & = & \left[n_F(\xi^{(+)}_{k} +l
\bar{\omega}_0) \right]
K(\xi^{(+)}_{k} +l \bar{\omega}_0 )  \nonumber \\
& + & \left[1-n_F(\xi^{(+)}_{k} -l \bar{\omega}_0) \right]
K(\xi^{(+)}_{k} -l \bar{\omega}_0 )
\end{eqnarray}
and $g_2(\xi^{(+)}_{{\bf k}}) $

\begin{equation}
g_2(\xi^{(+)}_{{\bf k}})= \left[n_F(\xi^{(+)}_{k} +\bar{\omega}_0)
\right] B(\xi^{(+)}_{{\bf k}} +\bar{\omega}_0 )+
\left[1-n_F(\xi^{(+)}_{k} -\bar{\omega}_0) \right] B(\xi^{(+)}_{k}
- \bar{\omega}_0 ),
 \label{49r}
\end{equation}
where $B(x)=2 \pi  [H(x)+H(-x)]$, with $H(x)$ given in
eq.(\ref{16re}). In the $CO$ phase the scattering rate has a gap
due to the dimerization and the process of phonon spontaneous
emission by the quasi-particles. \cite{perroni1}

The role of the scattering rate is important not only to improve
the approximations of calculation of the spectral properties, but
also the optical properties. Through the scattering rate, we can
consider the new Green's function $\tilde{G}_{ret}^{(co)}(k,t)$

\begin{equation}
\tilde{G}_{ret}^{(co)}(k,t) =u^2_{k} \tilde{G}_{ret}^{(+)}(k,t)+
v^2_{k} \tilde{G}_{ret}^{(-)}(k,t), \label{16arzo1}
\end{equation}
that substitutes the function $G_{ret}^{(co)}(k,t) $ given in
eq.(\ref{16arzo}). In eq.(\ref{16arzo1}) the Green' functions
$\tilde{G}_{ret}^{(\nu)}(k,t)$ are
\begin{equation}
\tilde{G}_{ret}^{(\nu)}(k,t) =-i \theta(t) \exp\left(-i
\xi_{k}^{(\nu)} t \right) \exp\left(-t \Gamma(k)/2 \right),
\label{16arzo2}
\end{equation}
with $\nu$ standing for $+$ or $-$. Therefore it is possible to
derive a new spectral function and density of states that include
lattice fluctuation effects beyond the $ZFA$ approach. A pseudogap
as precursor of the actual gap at stronger $el-ph$ couplings is
found in the density of states. \cite{perroni1}

The inclusion of the scattering rate is able to affect the
features of the infrared absorption. As in the $ZFA$ approach, the
conductivity is decomposed into a coherent and an incoherent term:
\begin{equation}
\Re \sigma_{Fluct}(\omega)= \Re \sigma_{Fluct}^{(coh)}(\omega) +
\Re \sigma_{Fluct}^{(incoh)}(\omega). \label{63raz}
\end{equation}
The coherent conductivity is given by

\begin{equation}
\Re \sigma_{Fluct}^{(coh)}(\omega)= \left( \frac{ 4 e^2
t^2}{\omega} \right) e^{ -2 S} \sum_{\nu_1,\nu_2} \int_{-\tilde{W}
}^{0} d \epsilon [n_F(\xi^{(\nu_1)}-\omega)-n_F(\xi^{(\nu_1)})]
\tilde{C}^{(\nu_1,\nu_2)}(\epsilon,\omega) h(\epsilon)
A^{(\nu_1,\nu_2)}(\epsilon), \label{130arz}
\end{equation}
where $\tilde{C}^{(\nu_1,\nu_2)}(\epsilon,\omega)$ is

\begin{equation}
\tilde{C}^{(\nu_1,\nu_2)}(\epsilon,\omega)= \frac{
\Gamma(\epsilon) } { \Gamma^2(\epsilon)+ \left( \xi^{(\nu_2)}
-\xi^{(\nu_1)} + \omega   \right)^2 },
 \label{131arz}
\end{equation}
$h(\epsilon)=g\left( \epsilon \right)\left( 1-\frac{\epsilon^2}{4
t^2_{eff} } \right)$, with $g(\epsilon)$ bare density of states,
and the function $A^{(\nu_1,\nu_2)}(\epsilon)$ is expressed by

\begin{equation}
A^{(+,+)}(\epsilon)=A^{(-,-)}(\epsilon)=
\frac{\epsilon^2}{\epsilon^2+E^2}
\end{equation}
with

\begin{equation}
A^{(+,-)}(\epsilon)=A^{(-,+)}(\epsilon)=
\frac{E^2}{\epsilon^2+E^2}.
\end{equation}

The latter term of the conductivity becomes

\begin{eqnarray}
\Re \sigma^{(incoh)}_{Fluct}(\omega)= && \left( \frac{ 2 e^2
t^2}{\omega} \right) e^{ -2 S_T} \sum_{\nu_1,\nu_2}
\int_{-\tilde{W} }^{0} d \epsilon \int_{-\tilde{W} }^{0} d
\epsilon_1 g(\epsilon) g(\epsilon_1)
R^{(\nu_1,\nu_2)}(\epsilon,\epsilon_1,\omega),
\label{134arz}
\end{eqnarray}
where the function $ R^{(\nu_1,\nu_2)}(\epsilon,\epsilon_1,\omega)
$ is given by

\begin{equation}
 R^{(\nu_1,\nu_2)}(\epsilon,\epsilon_1,\omega)=
\sum_{l=1}^{+ \infty} \frac{\left( 2 g^2 f^2 e^{-4 \alpha}
\right)^l }{l!}
\left[
J_l^{(\nu_1,\nu_2)}(\epsilon,\epsilon_1,\omega)+
H_l^{(\nu_1,\nu_2)}(\epsilon,\epsilon_1,\omega) \right],
\label{135ar}
\end{equation}
$C^{(\nu_1,\nu_2)}(\epsilon,\epsilon_1,x)$ is

\begin{equation}
 C^{(\nu_1,\nu_2)}(\epsilon,\epsilon_1,x)=
\frac{1}{2} \frac{ [\Gamma(\epsilon)+\Gamma(\epsilon_1)] } {
[\Gamma(\epsilon)+\Gamma(\epsilon_1)]^2/4+
(\xi^{(\nu_1)}-\xi_1^{(\nu_2)}+x)^2 } \label{131r}
\end{equation}
and

\begin{equation}
\xi_{i}^{(\nu_{j})}=\nu_{j} \sqrt{\epsilon^2_{i}+E^2}.
\end{equation}
In eq.(\ref{134arz}) the functions $J_l^{(\nu_1,\nu_2)}(
\epsilon,\epsilon_1,\omega )$

\begin{eqnarray}
J_l^{(\nu_1,\nu_2)}( \epsilon,\epsilon_1,\omega ) =
C^{(\nu_1,\nu_2)}( \epsilon,\epsilon_1,\omega+l\bar{\omega}_0 )
[n_F(\xi^{(\nu_2)}_1-l\bar{\omega}_0-\omega)-
n_F(\xi^{(\nu_2)}_1-l\bar{\omega}_0)] n_F(\xi^{(\nu_2)}_1)
\label{136ar}
\end{eqnarray}
and $H_l^{(\nu_1,\nu_2)}( \epsilon,\epsilon_1,\omega )$

\begin{eqnarray}
H_l^{(\nu_1,\nu_2)}( \epsilon,\epsilon_1,\omega ) &=&
C^{(\nu_1,\nu_2)}( \epsilon,\epsilon_1,\omega-l\bar{\omega}_0 )
[n_F(\xi^{(\nu_2)}_1+l\bar{\omega}_0-\omega)-
n_F(\xi^{(\nu_2)}_1+l\bar{\omega}_0)] \times
\nonumber \\
&& \times
\left[1-n_F(\xi^{(\nu_2)}_1) \right] \label{137ar}
\end{eqnarray}
describe phonon scattering processes.

In Fig.4 we show the conductivity at $t=10 \omega_0$ and
$\lambda=0.45$: the solid line represents the quantity obtained
adding fluctuations over $ZFA$, the dashed line that derived
within $ZFA$ and dotted line that of the mean-field approach. The
values of the parameters have been chosen such that they are
appropriate for quasi $1D$ inorganic metals. First we observe that
the inverse square-root singularity obtained in the mean-field and
$ZFA$ approach is reduced. Then, due to quantum lattice
fluctuations, the new conductivity not only shows a long subgap
tail but presents anomalies also over the gap when compared with
the mean-field spectrum. Therefore it strengthens the tendencies
already shown by the $ZFA$ conductivity and it is in good
agreement with the experimental spectra.
\cite{ross,degiorgi1,schwartz}

As stressed in the introduction, an analytical variational
approach, \cite{zhe1,wang} valid in the weak $el-ph$ coupling
regime and based on a similar procedure of calculation, has been
developed to study the optical conductivity. In order to emphasize
the different physical results between the present work and the
previous one, in Fig.5 we compare the conductivity $\Re
\sigma_{Fluct}$ (solid line) with the corresponding quantity
(dashed line) obtained within the preceding approach \cite{wang}
for $t=10 \omega_0$ and $\lambda=1$. This last conductivity is
extrapolated from the weak coupling limit and, as already noted by
the authors of the work, it shows a maximum that does not coincide
with their energy gap (see dashed arrow in the figure). This
feature is not consistent with experimental spectra
\cite{degiorgi1} that, at $T=0$, show the maximum of the infrared
optical conductivity generally near to the gap determined by other
experimental measurements such as the low-temperature resistivity,
the temperature dependence of the susceptibility and neutron and
Raman scattering. \cite{schwartz} As clearly shown in Fig.5, our
approach is not affected by this problem (the gap in our approach
is indicated by the solid arrow).  While in the previous work,
\cite{wang} the presence of a tail below the peak should be due to
usual interband transitions, in our approach the peak in the
conductivity corresponds to the gap in the quasi-particle spectrum
and is well above a real subgap tail determined by lattice
fluctuations.

In the inset of Fig.5 we report the ratio (solid line) between $S
(\omega)$, the spectral weight of the conductivity calculated in
this section, and $S_m$, the same quantity in the limit of
infinite frequency. This ratio is compared with that (dashed line)
obtained by means of exact diagonalizations of the Hamiltonian for
a system of six sites. \cite{weisse} Our approach captures the
correct onset of the optical response, even if it presents sharper
features. This could be due to the fact that we are not
considering a small cluster but we are performing the
thermodynamic limit, and, furthermore, that higher order
self-energy insertions should be included in order to improve the
approach.

The optical response shows similar features at lower values of the
adiabaticity ratio $t/\omega_0$. In the regime of near electronic
and phononic energy scales the effects due to lattice fluctuations
are enhanced and the contributions below and above the gap
subtract a larger spectral weight to the interband gap term. This
regime is typically important for inorganic linear chain
compounds,  for example, the compound $TTF-TCNQ$, that is a
narrow-band one-dimensional metal with a relatively strong $el-ph$
coupling. \cite{Heeger}

\section{Conclusions}
We have discussed the optical properties of the half-filled
spinless Holstein model in the $1D$ case at zero temperature
within the $ZFA$ approach. We have observed that, with increasing
the $el-ph$ coupling, the ordered phase affects the conductivity
spectra inducing a transfer of spectral weight from low to high
energies and a broadening of the optical interband gap. The
quantum lattice fluctuations considered in the $ZFA$ approach are
able to affect the optical response of the system that shows bands
below and above the gap. When fluctuation effects beyond the $ZFA$
approach are included, the optical conductivities are profoundly
changed since they are characterized by a subgap tail and a wide
band above the interband optical gap term that is smoothed when
compared with the mean-field result. Therefore the inelastic
scattering processes influence the low-frequency and
high-frequency features of the conductivity in agreement with
experimental spectra. \cite{degiorgi1,schwartz} Our results make
clear that a treatment of the lattice fluctuations beyond the
$ZFA$ approach is required to obtain a consistent agreement with
experimental data.

In this paper lattice fluctuation effects beyond the $ZFA$
approach are included calculating perturbatively the scattering
rate of the polarons that form the ordered state. A second-order
perturbation calculation on the $ZFA$ solution reproduces the
integrated spectral weights obtained by exact numerical approaches
suggesting that the employed approach can capture the infrared
response of the model. Finally we note that our approach is valid
in the infrared range of frequencies where the interband
absorption typically takes place. Thus it is not able to reveal
the structures attributed to collective excitation modes arising
from the $CDW$ condensate. \cite{gruner,gruner1,degiorgi}

\section*{Figure captions}
\begin{description}

\item  {F1}
The phase diagram of the system. The line separates the $CO$ from
the $A$ phase, that represents the disordered normal state. In the
inset, the transition line between the $CO$ and the disordered $A$
phase calculated in the mean-field (white squares), $ZFA$ (white
circles) and $DMRG$ approach (black squares).

\item  {F2}
The renormalized density of states derived in the $ZFA$ (solid
line) and mean-field approach (dashed line)  at $t=5 \omega_0$ and
$g=2.2$ as a function of the energy (in units of $\omega_0$).

\item  {F3}
(a) The conductivity at $t=5 \omega_0$ and $g=2.4 $ as a function
of the frequency (in units of $\omega_0$) decomposed into its
different components: coherent term (dashed line), first
incoherent term (dotted line) and second incoherent term
(dash-dotted line).

(b) The conductivity spectra derived in the $ZFA$ (solid line) and
mean-field approach (dashed line) at $t=5 \omega_0$ and $g=2.4$ as
a function of the frequency (in units of $\omega_0$).

The conductivities are expressed in units of $\frac{e^2 \rho}{m t}$, with
$m= \frac {1}{2t}$.

\item  {F4}
The conductivity (in units of $\frac{e^2 \rho}{m t}$, with $m=
\frac {1}{2t}$) obtained including fluctuations beyond the $ZFA$
(solid line), derived in the $ZFA$ (dashed line) and in the
mean-field approach (dotted line) up to 14 $\omega_0$ at $t=10
\omega_0$ and $\lambda=0.45$.

\item  {F5}
The conductivity of the present approach (solid line) in
comparison with that (dashed line) calculated in a previous paper
\cite{wang} at $t=10 \omega_0$ and $\lambda=1.0$. The solid arrow
indicates the gap in our approach and the dashed arrow that
obtained in a previous work. \cite{wang} In the inset the ratio
between the spectral weight $S (\omega)$ and $S_m$ calculated in
the present approach (solid line) compared with the same quantity
(dashed line) derived from exact numerical diagonalizations.
\cite{weisse}

\end{description}

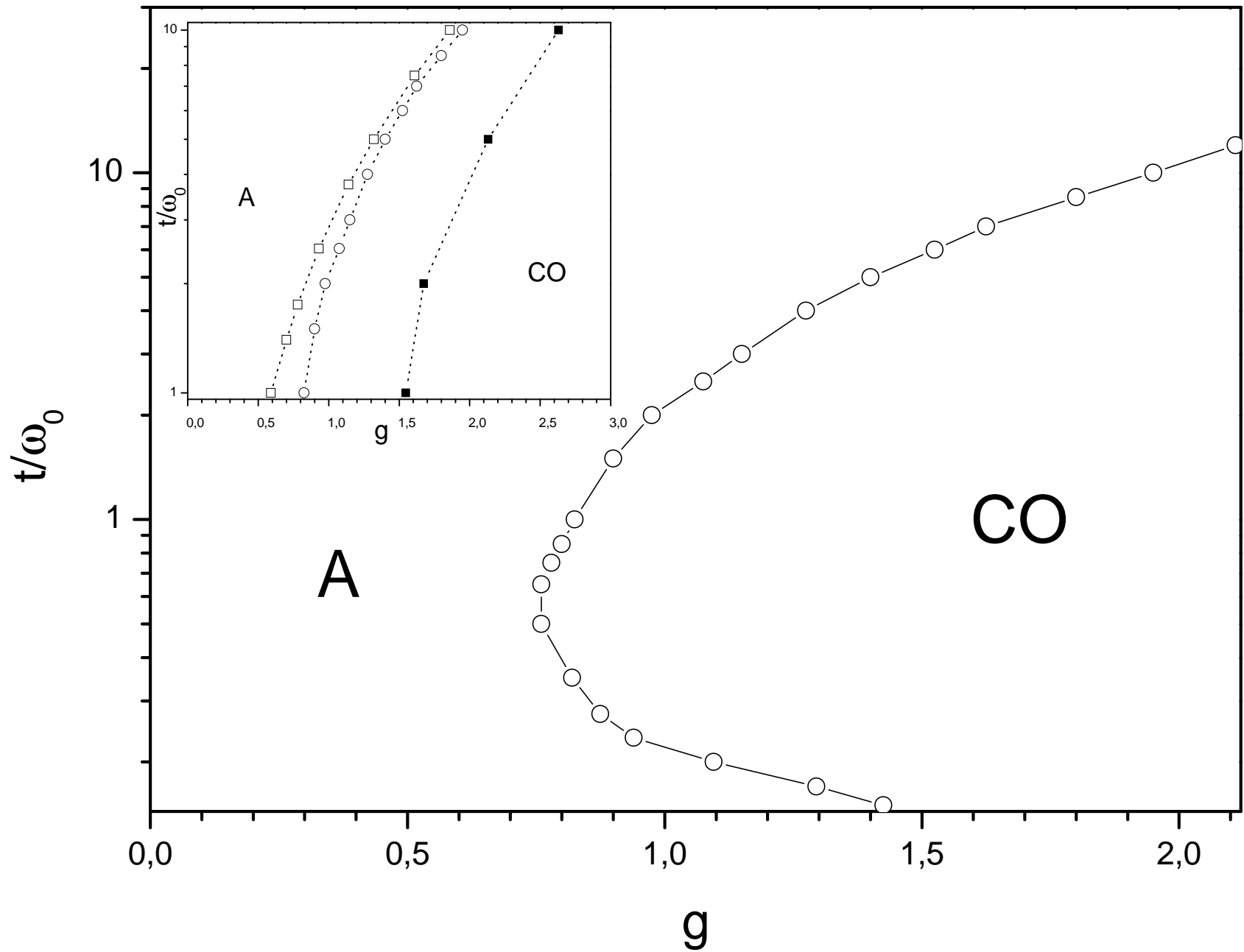
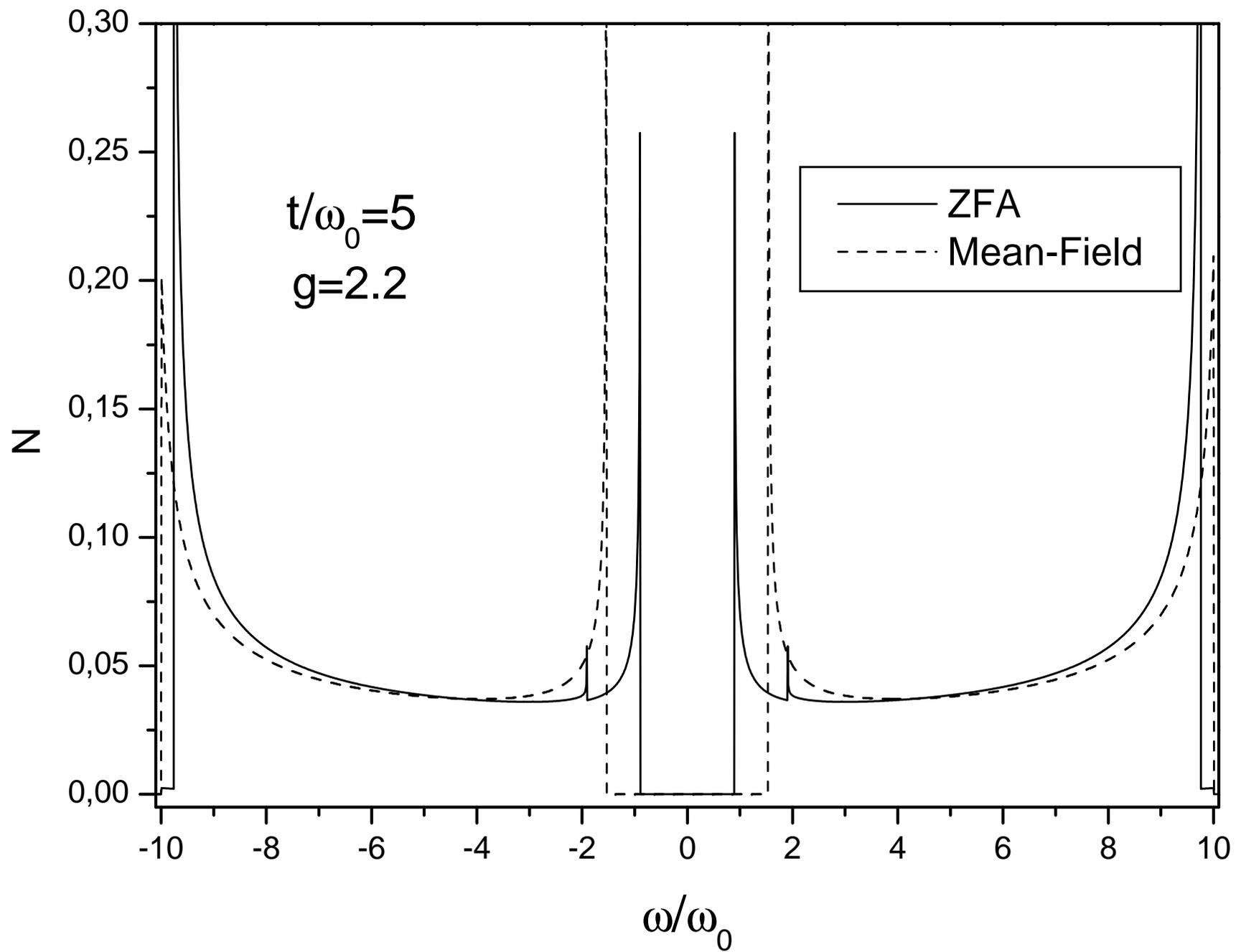
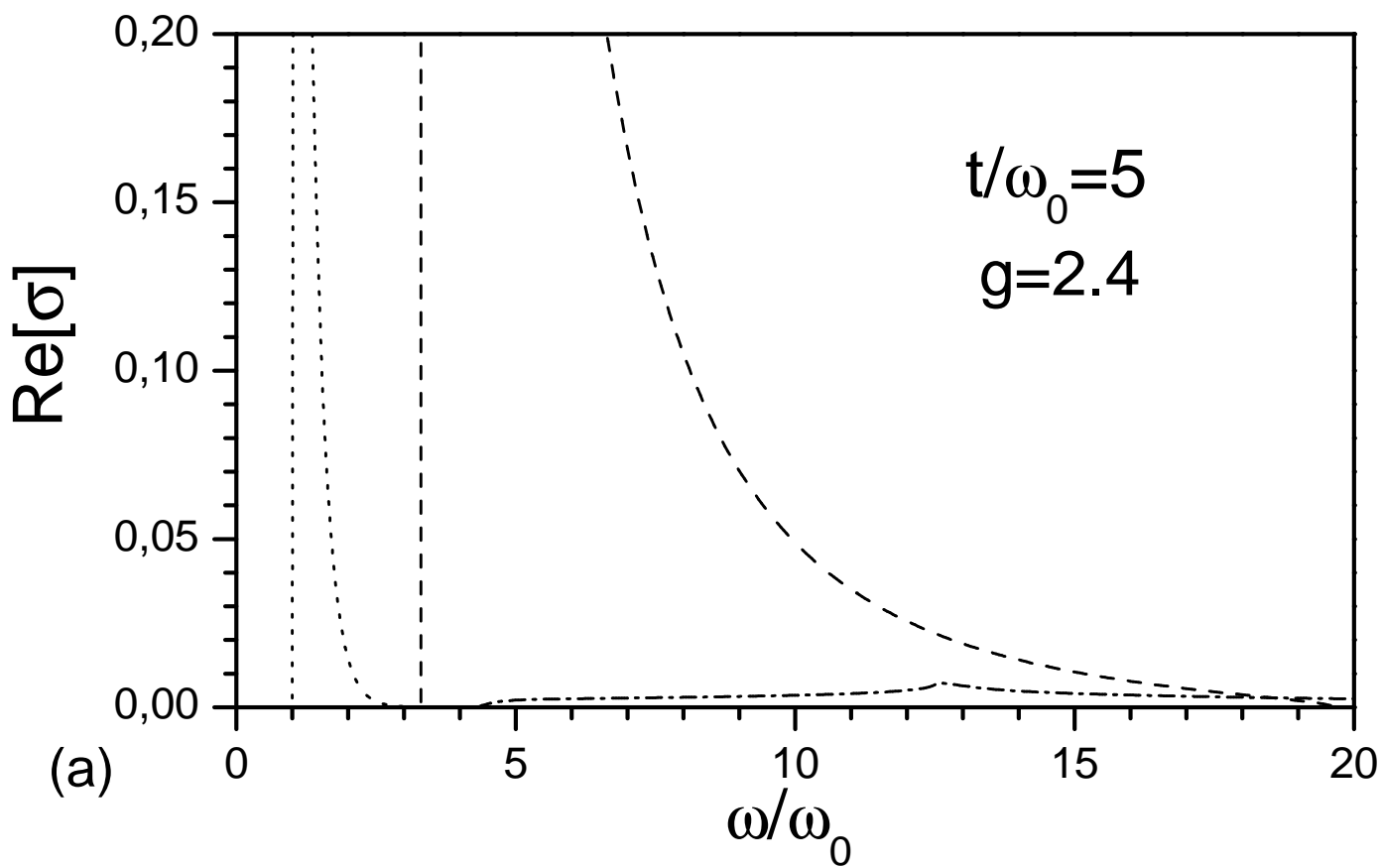
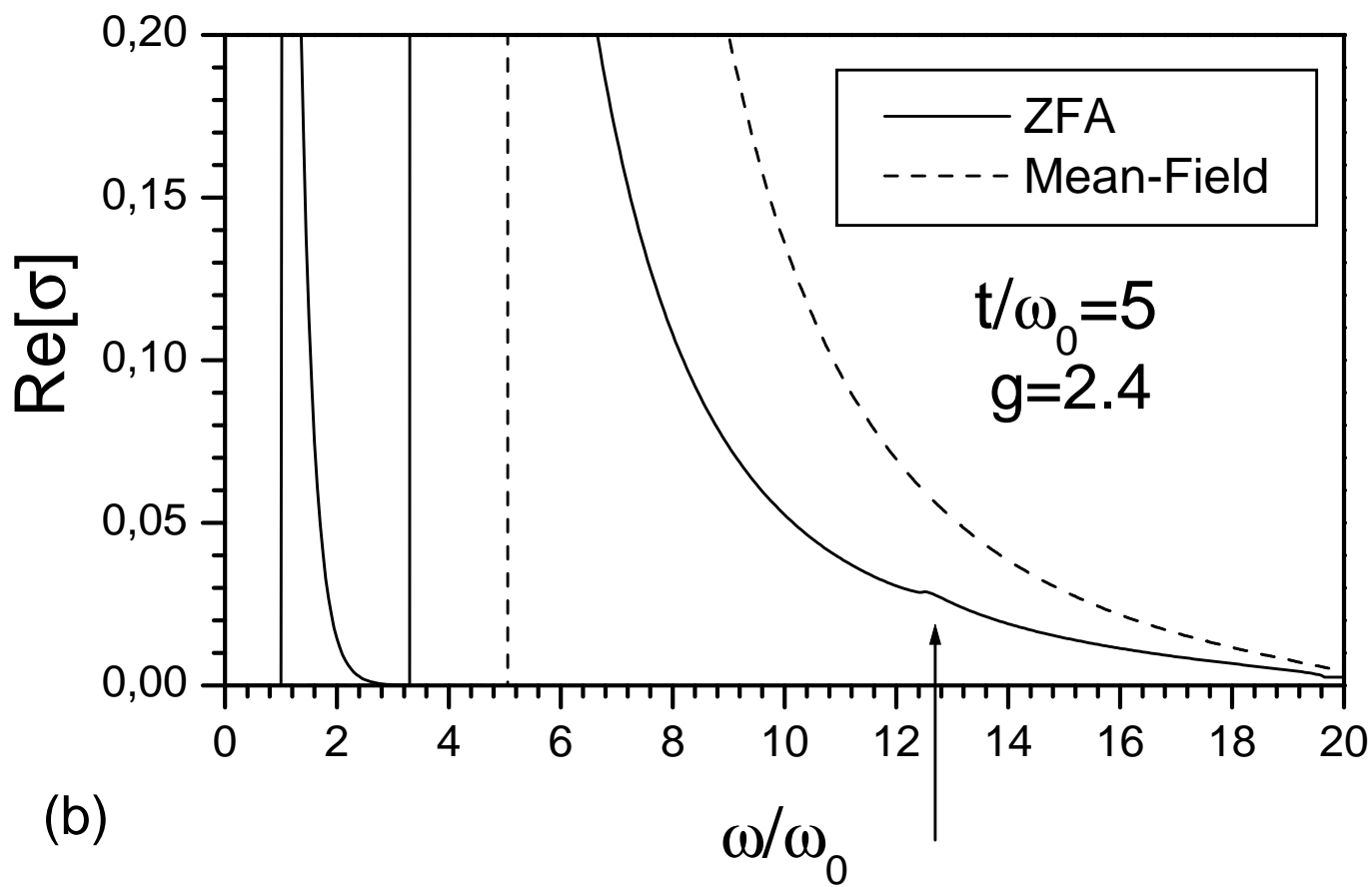
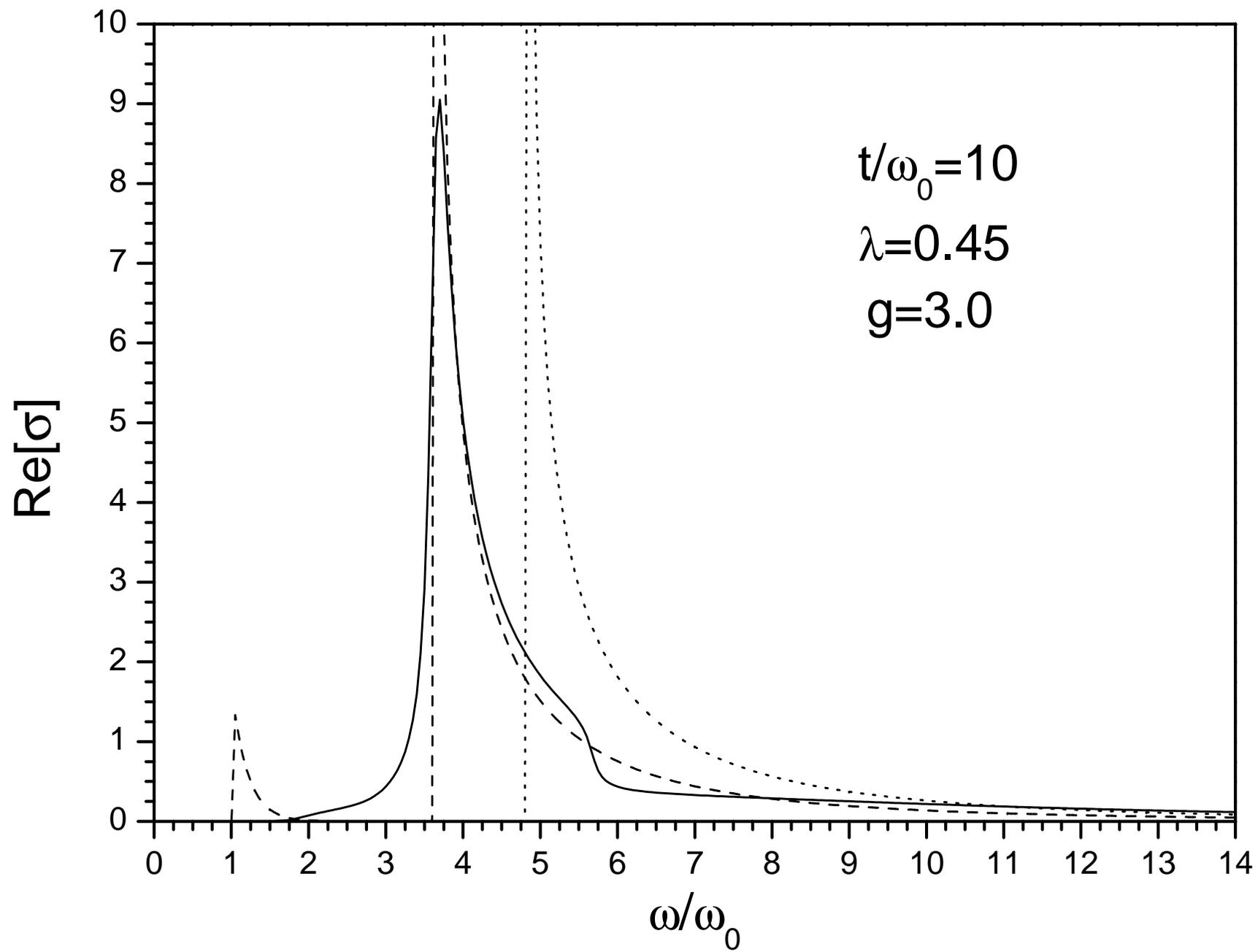
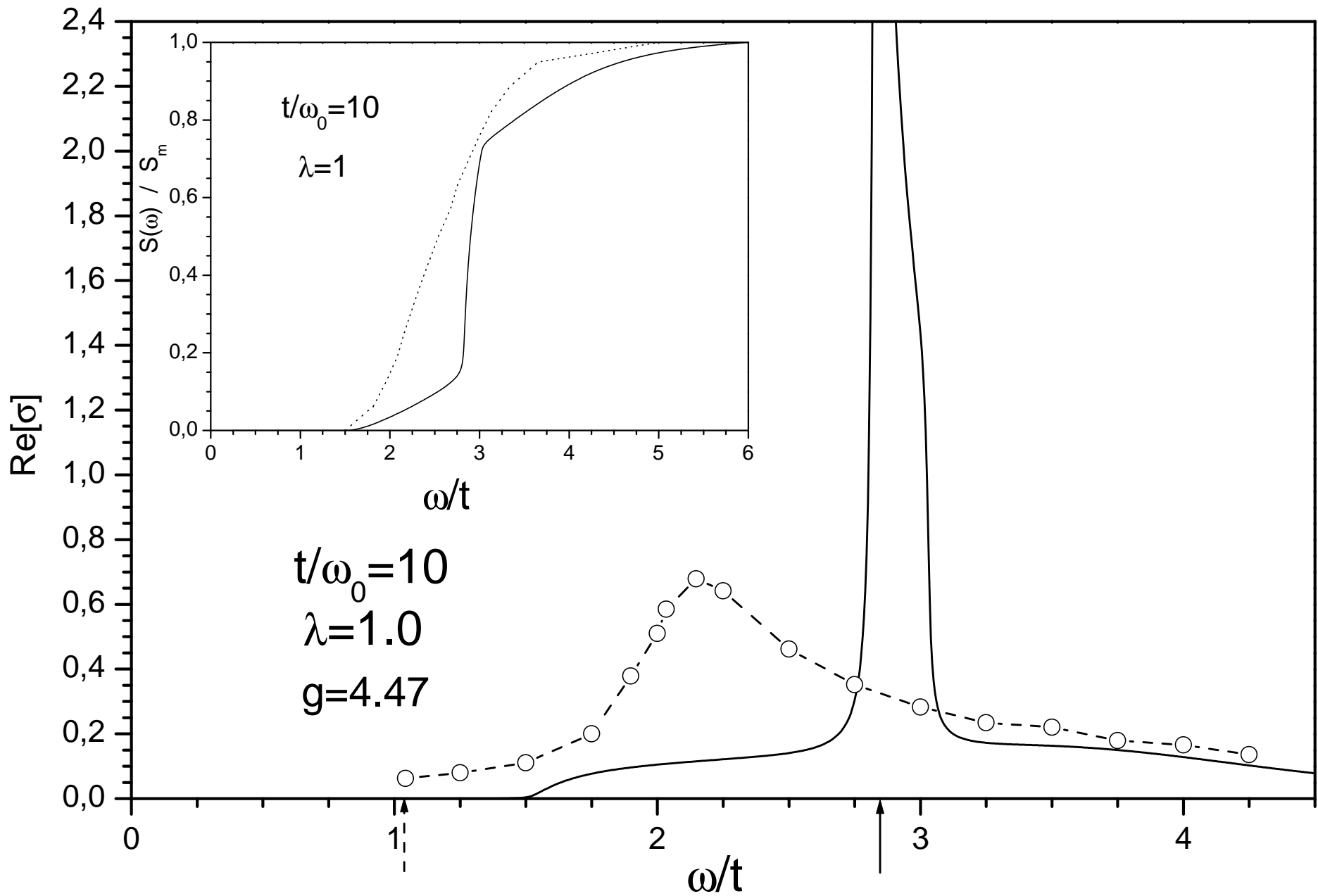
\end{document}